\documentclass[
    ,final            
  ]
  {aipproc}

\layoutstyle{6x9}

\usepackage{amsmath}
\usepackage{placeins}
\usepackage{wrapfig}
\usepackage{verbatim}
\usepackage{amssymb}

\newcommand{\ktsqav}{\langle k_{T}^{2}\rangle}
\newcommand{\ktsqavgT}{\langle k_{T}^{2}\rangle_{_{\rm WG}}} 

\begin{document}

\title{Impact of double TMD effects on transversity measurements at RHIC}



\author{Wilco J. den Dunnen\footnote{Speaker at the XIX International Workshop on Deep-Inelastic Scattering and Related Subjects \mbox{(DIS 2011)}, April 11-15, 2011, Newport News, VA USA} }{
	address={Department of Physics and Astronomy, 
Vrije Universiteit Amsterdam, 
De Boelelaan 1081, \mbox{1081 HV} Amsterdam, The Netherlands}
}

\author{Dani\"el Boer}{
	address={Theory Group, KVI, University of Groningen,
Zernikelaan 25, 9747 AA Groningen, \mbox{The Netherlands}}
}

\author{Aram Kotzinian}{
	address={Dipartimento di Fisica Teorica, Universit\`a di Torino
and \it INFN, Sezione di Torino, Via P. Giuria 1, I-10125 Torino, Italy
and \it Yerevan Physics Institute, 2 Alikhanyan Brothers ST., 375036 Yerevan, Armenia}
}

\begin{abstract}
A quantitative estimate is presented for the double transverse spin asymmetries \emph{at measured $q_T$} in both the Drell-Yan process and $W$-boson production due to Transverse Momentum Dependent (TMD) effects.
These spin asymmetries are calculated as a function of the lepton azimuthal angle as measured \emph{in the laboratory frame.}
In this frame, in contrast to the Collins-Soper frame, the TMD effects contribute to the spin asymmetry $A_{TT}(q_T)$ in the same way as transversity does, which makes them a background for transversity measurements in the Drell-Yan process and new physics studies in $W$-boson production.
Using the current knowledge of the relevant TMDs we conclude that this background is negligible and, therefore, will not hamper transversity measurements nor new physics studies when performed in the laboratory frame. 
We also point out a cross-check asymmetry measurement to bound the TMD contributions, which is independent of assumptions on the sizes of the relevant TMDs.
\end{abstract}

\addtolength{\topmargin}{-0.5cm}
\addtolength{\textheight}{1cm}
\maketitle


\section{Introduction}
Transversity was first discussed by Ralston and Soper \cite{Ralston:1979ys}, who suggested its measurement in the Drell-Yan (DY) process through the double transverse spin asymmetry $A_{TT}$ integrated over the transverse momentum $q_T$ of the lepton pair and at measured $q_T$ $A_{TT}(q_T)$, in particular at $q_T=0$.
At measured $q_T$ there will be background contributions from transverse momentum dependence of partons, that have not yet been considered. 
The $A_{TT}(q_T)$ asymmetry was estimated to be at most 5\% \cite{Kawamura:2007ze},
based on the upper bound on the transversity distribution.
The first extraction \cite{Anselmino:2007fs} indicates the quark transversity to be only half of its maximum value, which, if it also holds for the antiquarks, reduces the asymmetry to 1\%, making a background study relevant.
We will also study $W$-boson production, in which one expects zero contribution from transversity within the Standard Model \cite{Bourrely:1994sc,Rykov:1999ru}. This allows for new physics studies as proposed in \cite{Boer:2010mc}, where the maximal asymmetry was estimated to be around 1\%, reinforcing the need for a background study.

The two relevant TMD effects are the double Sivers effect, which describes a transverse momentum distribution of quarks inside a transversely polarized hadron which is asymmetric w.r.t.\ the spin direction \cite{Sivers:1989cc} and another effect that was first discussed by Ralston and Soper \cite{Ralston:1979ys}, which describes the distribution of longitudinally polarized quarks inside a transversely polarized hadron.
Both effects are described by a transverse momentum dependent parton distribution (TMD): the Sivers effect by a TMD often denoted by $f_{1T}^\perp$ \cite{Boer:1997nt} and the other by $g_{1T}$ \cite{Tangerman:1994eh} also referred to as Worm-Gear (WG) function.

The expressions for the double Sivers and double WG effect for DY 
have been given in Ref.\ \cite{Boer:1999mm,Lu:2007ev,Arnold:2008kf}.
One can consider the so-called Collins-Soper (CS) frame, which allows one to distinguish the double transverse spin asymmetries arising from transversity, the Sivers effect and the WG effect by their lepton azimuthal angular dependence.
However, in the laboratory frame, all three effects will contribute to the same angular distribution.
The lab frame is thus theoretically not the preferred frame to extract the transversity distribution, but it is experimentally more `direct' to do so.
In fact, in $W$-boson production with a leptonic decay, it is virtually impossible to transform to the CS frame due to the unobserved neutrino.
This makes the lab frame experimentally the most desirable frame to measure spin asymmetries.
To study the impact of the Sivers and WG effect on such a measurement, we will present quantitatively the size of the spin asymmetries \emph{in the lab frame} caused by partonic transverse momentum effects in both the DY process and $W$-boson production.

\section{Distribution functions}

A factorization between $k_T$ and $x$ dependence and a Gaussian dependence on $k_T$ will be assumed, i.e. we use for the unpolarized parton distribution
\begin{equation}
f_1^q(x,k_T) = f_1^q(x)\, \frac{1}{\pi \ktsqav} e^{-k_T^2/\ktsqav},
\end{equation}
with the value of the width $\ktsqav = 0.25\ \text{GeV}^2$, fitted by \cite{Anselmino:2005nn}. For the Sivers function, we will use the extraction obtained by \cite{Anselmino:2008sga}.
A determination of the Worm-Gear distribution based on fits of experimental data is not available, so we will employ a model for this WG function. We will use a Gaussian Ansatz in terms of its first transverse moment, i.e.\
\begin{equation}
g_{1T}^q(x_,k_T) = g_{1T}^{q(1)}(x)\, \frac{2 M_p^2}{\pi\ktsqavgT^2} 			e^{-k_T^2/\ktsqavgT}.
\end{equation}
For the width we will take a value in accordance with the bag model \cite{Avakian:2010br} $\ktsqavgT=0.71\ktsqav$ and for the first moment, we will use a Wandzura-Wilczek type approximation \cite{Wandzura:1977qf,Tangerman:1994bb,Kotzinian:1995cz} to express it in terms of the known helicity distribution $g_1(x)$ by
\begin{equation}
g_{1T}^{q(1)}(x) \approx x \int_x^1 dy\, \frac{g_1^q(y)}{y}.
\end{equation}
The resulting functions agree with model calculations and lattice evaluations, see \cite{Boer:2011vq} for details.
For numerical estimations the DSSV helicity distribution \cite{deFlorian:2008mr} is used.

\section{Spin asymmetries in the Drell-Yan process}

\begin{figure}[h]
\includegraphics[height=3.3cm]{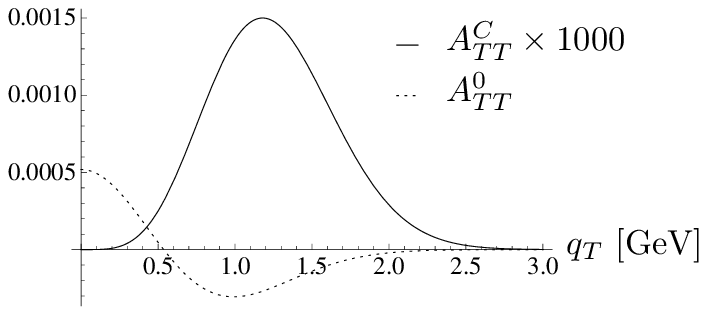}
\includegraphics[height=3.3cm]{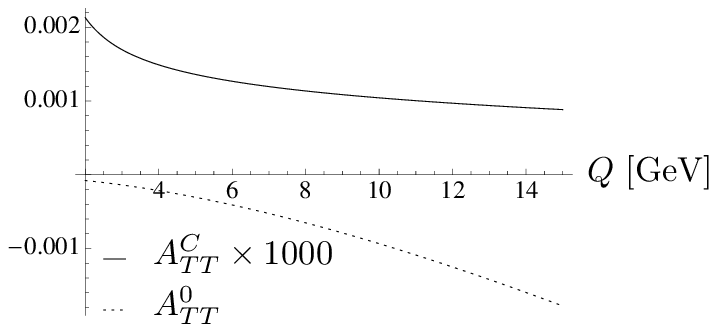}
\caption{Contribution to $A_{TT}(q_T,Q,Y)$ in the Drell-Yan process from the double Sivers effect at $\sqrt{s}=500$ GeV as a function of $q_T$ at $Q=5$ GeV (left) and $Q$ at $q_T=1$ GeV (right) valid for $|Y| \lesssim 2$.}
\label{ATTDY1}
\end{figure}

We will define a spin flip symmetric and antisymmetric cross section as a function of the transverse momentum $q_T$, total momentum $Q$ and rapidity $Y$ of the lepton pair and the lepton azimuthal angle $\phi_l$ (measured w.r.t.\ the spin plane) in the laboratory frame by
\begin{equation}\label{CSinWL}
d\sigma^{S,A}(q_T,Q,Y,\phi_l)	\equiv \frac{1}{4}\left(
		d\sigma^{\uparrow\uparrow} \pm d\sigma^{\uparrow\downarrow} 
		\pm d\sigma^{\downarrow\uparrow} + 
		d\sigma^{\downarrow\downarrow} \right).\\
\end{equation}
Two double transverse spin asymmetries will be defined,
\begin{equation}\label{ATTdef}
\begin{aligned}
A_{TT}^0(q_T,Q,Y)	&\equiv \int_0^{2\pi}d\phi_l \, d\sigma^A \Bigg/
		\int_0^{2\pi}d\phi_l \, d\sigma^S,\\
A_{TT}^C(q_T,Q,Y) 	&\equiv \left(\int_{-\pi/4}^{\pi/4} -
  		\int_{\pi/4}^{3\pi/4} + \int_{3\pi/4}^{5\pi/4} -
  		\int_{5\pi/4}^{7\pi/4}\right)d\phi_l \, d\sigma^A \Bigg/
  		\int_0^{2\pi}d\phi_l \, d\sigma^S,
\end{aligned}
\end{equation}
which select out the $\phi_l$ independent part of the cross section and the part $\propto \cos2\phi_l$. The double Sivers effect contribution to both these asymmetries is plotted in Fig. \ref{ATTDY1}, the share coming from the WG effect is left out being negligible compared to this. For detailed expressions we refer to \cite{Boer:2011vq}.

The $A_{TT}^0$ asymmetry reaches up to the percent level, but only for large $Q^2$ outside the range of interest, whereas the $A_{TT}^C$ asymmetry receives a contribution at the level of $10^{-6}$ from the double Sivers effect and $10^{-8}$ from the double WG effect.
The maximal value of $A_{TT}^C$ is bounded by the maximal value of $A_{TT}^0$, irrespective of the parameterization used for TMD distributions.
Therefore, as a cross-check of the smallness of the TMD background, one can verify that the $A_{TT}^0$ asymmetry is indeed much smaller.
\\


\FloatBarrier
\section{Spin asymmetries in $W$-boson production}
\begin{figure}[htb]
\includegraphics[height=4cm]{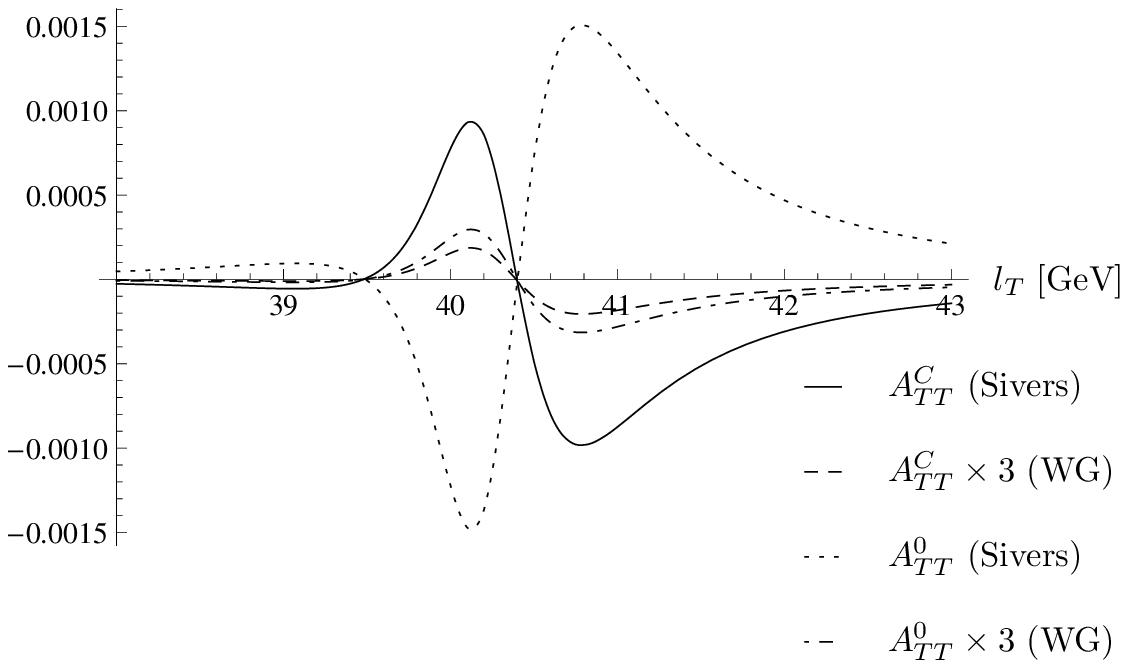}
\hspace{1cm}
\includegraphics[height=4cm]{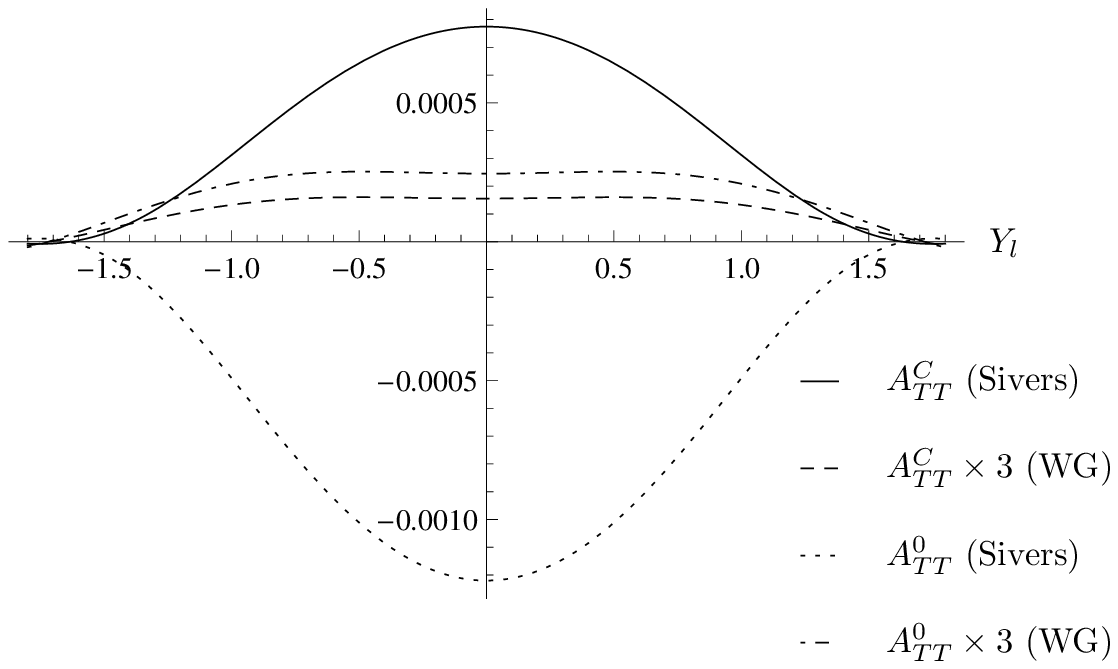}
\caption{Contributions to $A_{TT}(l_T,Y_l)$ in $W^+$ boson production from the double Sivers and Worm-Gear effect at $\sqrt{s}=500$ GeV as a function of $l_T$ at $Y_l=0$ (left) and $Y_l$ at $l_T=40$ GeV (right).}\label{ATTWp}
\end{figure}

In $W$-boson production we define the same asymmetries as in Eq. \ref{ATTdef}, but we anticipate on the neutrino being unobserved and express the asymmetries as a function of the lepton transverse momentum $l_T$ and rapidity $Y_l$ only.
We show the asymmetries in $W^+$ production in Fig.\ \ref{ATTWp}, because they are largest.
The maximal asymmetry is near resonance and reaches up to $0.15\%$, which is already below the detection limit at RHIC.
However, for a bound on a possible $W$-$W^\prime$ mixing it is the asymmetry in the integrated cross section that is relevant. In those asymmetries the contribution at $l_T<M_W/2$ largely cancels the contribution at $l_T>M_W/2$, resulting in very small asymmetries. 
We find the asymmetry in the integrated cross section in $W^\pm$ production below $10^{-6}$, forming a negligible background for the studies proposed in \cite{Boer:2010mc}.

\FloatBarrier

\section{Conclusions}
We estimated the contribution from the Sivers and Worm-Gear effect to the double spin asymmetries at measured $q_T$ in the DY process and in $W$-boson production. Our conclusion is that, in the laboratory frame, these TMD effects contribute to the lepton azimuthal angle dependent $A_{TT}^C$ asymmetry, but only at the level of $10^{-6}$ in the DY process and $10^{-3}$ in $W$-boson production. At that level, the TMD effects do not form a relevant background for transversity measurements nor for new physics studies based on $A_{TT}^C$ in the DY process and $W$-boson production respectively. As a cross-check one can use the azimuthal angle independent $A_{TT}^0$ asymmetry to bound the TMD contributions.



\bibliographystyle{aipproc}   

\end{document}